\documentclass[twocolumn,prl,unsortedaddress,showpacs,floatfix,citeautoscript,nofootinbib]{revtex4}
\usepackage{amssymb}
\usepackage{amsmath}
\usepackage{amsbsy}
\usepackage{latexsym,epsfig,graphicx}
\usepackage{dcolumn}
\usepackage{bm}
\usepackage[normalem]{ulem}
\usepackage{graphicx}
\usepackage{subfigure}
\usepackage{color}
\usepackage[colorlinks,urlcolor=blue,citecolor=blue]{hyperref}

\begin{document}

\title{Magnetic Solitons in a Binary Bose-Einstein Condensate}
\author{Chunlei Qu$^{1}$}
\author{Lev P. Pitaevskii$^{1,2}$}
\author{Sandro Stringari$^{1}$}
\affiliation{$^{1}$INO-CNR BEC Center and Dipartimento di Fisica, Universit\`a di Trento,
38123 Povo, Italy \\
$^{2}$Kapitza Institute for Physical Problems RAS, Kosygina 2, 119334
Moscow, Russia}
\date{\today }

\begin{abstract}
We study solitary waves of polarization (magnetic solitons) in a
two-component Bose gas with slightly unequal repulsive intra- and interspin
interactions. In experimentally relevant conditions we obtain an analytical
solution which reveals that the width and the velocity of magnetic solitons
are explicitly related to the spin healing length and the spin sound
velocity of the Bose mixture, respectively. We calculate the profiles, the
energy and the effective mass of the solitons in the absence of external
fields and investigate their oscillation in a harmonic trap where the
oscillation period is calculated as a function of the oscillation amplitude.
The stability of magnetic solitons in two dimensions and the conditions for their
experimental observation are also briefly discussed.
\end{abstract}

\maketitle


\textit{Introduction.---}Solitons, the fascinating topological excitations
of nonlinear systems, have drawn a considerable research interest in many
physical branches ranging from classical fluids, fibre optics~\cite%
{Mollenauer1980}, polyacetylene~\cite{Su1980}, magnets~\cite{Kosevich1990}
and so on. Because of the interplay of nonlinearity and dispersion, solitons can
move in their medium without loosing their shape and thus have important
application in information processing. Among various physical systems,
ultracold atomic gases provide a prominent platform for the investigation of
solitons which can be engineered by phase imprinting, density imprinting,
quantum quenches, etc. Soon after the realization of Bose-Einstein
condensation, dark and bight solitons characterized by density notches and
density bumps have been actually observed in repulsive~\cite%
{Burger1999,Denschlag2000} and attractive~\cite{Khaykovich2002} interacting
Bose gases, respectively.

Recently, vector solitons such as dark-dark and dark-bright soliton
complexes have been explored in spinor Bose gases where the underlying
physics is even richer (see, for example, reference~\cite{Busch2001}).
Different techniques have been utilized to generate vector solitons in
experiments with quantum mixtures. For instance, by filling the dark soliton
of one species with atoms of another species~\cite{Becker2008} or by the
counterflow of two superfluids~\cite{Hamner2011}, the dark-bright solitons
are engineered and observed. Despite this experimental progress, our
theoretical understanding of vector solitons mainly relies on numerical
calculations and some analytical results were only obtained under stringent
conditions. For example, Busch and Anglin~\cite{Busch2001} studied the
dark-bright soliton under the assumption of equal spin interactions, $%
g_{11}=g_{22}=g_{12}$, where a dark soliton is developed in one component
and filled by the atoms of another component, the density background being
fully polarized (see also \cite{Kamchatnov2015} and references therein.)

In this Letter, we investigate another type of soliton, a \textit{magnetic
soliton}, in a two-component Bose gas which exists only when the repulsive
intra- and interspin interactions of the two species are unequal. Different
from dark-bright solitons, the magnetic soliton manifests itself as a
localized spin polarization $n_1-n_2$, where $n_{1,2}$ are the densities of
the two components, and resides in a spin-balanced density background. To
construct an explicit analytic solution, we take advantage of the fact that
typical experimental mixtures of hyperfine states of bosonic alkali atoms
are near the boundary of phase separation instability, that is the coupling
constants satisfy the inequality: 
\begin{equation}
\delta g\equiv g-g_{12}\ll g,  
\label{main}
\end{equation}
where $g=\sqrt{g_{11}g_{22}}$ and $\delta g>0$ in order to avoid phase
separation. For instance, considering the two hyperfine states $|F=1;
m_F=\pm 1\rangle$ of $^{23}$Na, one has $g_{11}=g_{22}$ and $\delta
g/g\approx 0.07$ \cite{Knoop2011,RbK}. The inequality (\ref{main}) is
crucial in order to ensure the decoupling between density and spin dynamics.
In particular it ensures that the total density $n=n_{1}+n_{2}$ is
practically unperturbed in the region of the magnetic soliton and one can
safely assume $n=const$ (see discussion below). A
theory of weakly-nonlinear polarization waves in two-component Bose gases
was developed in~\cite{Kamchatnov2014} beyond the condition of Eq.~(\ref{main}).

Similarly to a regular dark soliton, the magnetic soliton is expected to
show snake instability when the transverse size becomes larger than the
width of the soliton, due to its negative effective mass. However, since the
spin healing length is large in spinor Bose gases under the condition (\ref%
{main}), the magnetic solitons are more resilient against instability than
dark solitons whose width is fixed by the smaller density healing length.
For the same reason they can be wide enough to be observed directly in 
\textit{in situ} measurements. Finally, we calculate the oscillation
frequency of the magnetic solitons with arbitrary oscillation amplitudes in
a harmonic trap which may be a useful benchmark for the observation in real
experiments.

\textit{Densities and phases.---}We consider a two-component Bose-Einstein
condensate at $T=0$. The system is governed by two coupled Gross-Pitaevskii
equations (GPE) which can be obtained from the Lagrangian density $\mathcal{L} =\sum_{j=1,2}\frac{i\hbar }{2}\left( \psi _{j}^{\ast }\partial _{t}\psi _{j}-\psi _{j}\partial _{t}\psi
_{j}^{\ast }\right)-\mathcal{E}$, where $\psi _{j}$ is the condensate wave function for the $j$th component and the energy density is given by
\begin{equation}
\mathcal{E}=\sum_{j}\bigg[ \frac{\hbar ^{2}}{2m}|\nabla \psi _{j}|^{2}
+V_{ext}|\psi _{j}|^{2}+\sum_l\frac{g_{jl}}{2}|\psi
_{j}|^{2}|\psi _{l}|^{2}\bigg],  \label{energy}
\end{equation}
with $V_{ext}$ the external trapping potential. For simplicity, we will assume $g_{11}=g_{22}=g$ in the following.

We will first study magnetic solitons in the absence of the external
potential ($V_{ext}=0$). For a 1D two-component Bose gas the wave function
can be parametrized as
\begin{equation}
\left( 
\begin{array}{c}
\psi _{1} \\ 
\psi _{2}%
\end{array}%
\right) =\sqrt{n}\left( 
\begin{array}{c}
\cos (\theta /2)e^{i\varphi _{1}} \\ 
\sin (\theta /2)e^{i\varphi _{2}}%
\end{array}%
\right) ,
\end{equation}%
where $\varphi_{1,2}$ are the phases of the two components, $n$ is the 1D constant total density and the spin polarization is given by $\cos\theta=(n_1-n_2)/n$. It is also convenient to introduce the relative
phase $\varphi _{A}=\varphi_{1}-\varphi_{2}$ and the total phase $\varphi
_{B}=\varphi _{1}+\varphi_{2}$ of the two order parameters. Without any loss
of generality we can assume $\varphi_{1,2}=0$ at $z=-\infty$. In terms of
these new variables, the Lagrangian can be rewritten as 
\begin{eqnarray}
\mathcal{L}&=&-\frac{n\hbar }{2}\left( \cos \theta \partial _{t}\varphi
_{A}+\partial _{t}\varphi _{B}\right) -\frac{n\hbar ^{2}}{8m}\bigg[2\cos
\theta \partial _{z}\varphi _{A}\partial _{z}\varphi _{B}  \notag \\
&+&\left( \partial _{z}\varphi _{A}\right) ^{2}+\left( \partial _{z}\varphi
_{B}\right) ^{2}+\left( \partial _{z}\theta \right) ^{2}\bigg]+\frac{%
n^{2}\delta g}{4}\sin ^{2}\theta ,  \label{Eq-Lagrangian2}
\end{eqnarray}%
which represents a special case of the Lagrangian derived by Son and
Stephanov in Ref.~\cite{Son2002} where an additional coherent Rabi coupling was
considered~\cite{note}. It is important to note that the term $\partial
_{t}\varphi _{B}$, as a derivative, does not contribute to equations of
motion and will be omitted.

To look for traveling soliton solutions, we substitute $\theta =\theta(z-Vt) 
$ and $\varphi _{A,B}=\varphi _{A,B}(z-Vt)$ in the Lagrangian which is
simplified to the following form 
\begin{eqnarray}
\frac{\mathcal{L}}{nmc_{s}^{2}} &=&U(\cos \theta \partial _{\zeta }\varphi
_{A})-\frac{1}{2}[2\cos \theta \partial _{\zeta }\varphi _{A}\partial
_{\zeta }\varphi _{B}  \notag \\
&+&(\partial _{\zeta }\varphi _{A})^{2}+(\partial _{\zeta }\varphi
_{B})^{2}+(\partial _{\zeta }\theta )^{2}]+\frac{1}{2}\sin ^{2}\theta ,
\label{Eq-Lagrangin}
\end{eqnarray}%
with $\zeta =(z-Vt)/\xi _{s}$, $U=V/c_{s}$ as the coordinate and velocity in
unit of the spin healing length $\xi _{s}=\hbar /\sqrt{2mn\delta g}$ and spin sound velocity
$c_{s}=\sqrt{n\delta g/2m}$, respectively.

The variation of the Lagrangian with respect to the total phase $\varphi_{B} 
$ gives $\partial _{\zeta }\left( \partial_{\zeta }\varphi _{B}+\cos\theta
\partial _{\zeta }\varphi _{A}\right) =0$. Imposing the boundary condition $%
\partial_\zeta\varphi_{A,B}=0$ at $\zeta=\pm\infty$, we can write 
\begin{equation}
\partial _{\zeta }\varphi _{B}+\cos \theta \partial _{\zeta }\varphi _{A}=0.
\label{Eq-totalphase}
\end{equation}%
Therefore the total phase $\varphi _{B}$ can be calculated by a simple
integration once the other variables $\theta$ and $\varphi_A$ are
determined. With the help of Eq.~(\ref{Eq-totalphase}), the Lagrangian can
be further simplified and the variation with respect to $\varphi _{A}$ and $%
\theta $ gives the following two differential equations 
\begin{eqnarray}
\partial _{\zeta }\varphi _{A} = U\frac{\cos \theta }{\sin ^{2}\theta },
\qquad \partial _{\zeta }^{2}\theta = U^{2}\frac{\cos
\theta }{\sin ^{3}\theta }-\sin \theta \cos \theta ,  \label{Eq-theta}
\end{eqnarray}%
where we have used the additional boundary condition $n_{1,2}=n/2$, i.e, $%
\theta= \pi/2$, at $\zeta =\pm \infty$. The density distributions of the two
components can be obtained after a simple integration of the second Eq.~(\ref{Eq-theta}%
) and take the form 
\begin{equation}
n_{1,2} =\frac{n}{2}(1\pm\cos \theta )=\frac{n}{2}\left[ 1\pm \frac{\sqrt{%
1-U^{2}}}{\cosh (\zeta \sqrt{1-U^{2}})}\right].  \label{Eq-n1}
\end{equation}%
Without loss of generality, we will assume $n_1\geq n_2$ in the following discussions and
take $+$ sign in Eq.~(\ref{Eq-n1}).
Substituting result (\ref{Eq-n1}) for $\theta$ into the first Eq.~(\ref%
{Eq-theta}) and Eq.~(\ref{Eq-totalphase}), the relative phase $%
\varphi _{A}$ and the total phase $\varphi _{B}$ can be readily solved: 
\begin{eqnarray}
\cot \varphi_A &=&- \sinh(\zeta\sqrt{1-U^2})/U,  \label{Eq-phiA} \\
\tan (\varphi_B+C) &=& -\sqrt{1-U^2}\tanh(\zeta\sqrt{1-U^2})/U,  \label{Eq-phiB}
\end{eqnarray}
where the constant $C$ can be chosen to ensure $\varphi_B(\zeta=-\infty)=0$.

\begin{figure}[b!]
\includegraphics[width=8.0cm]{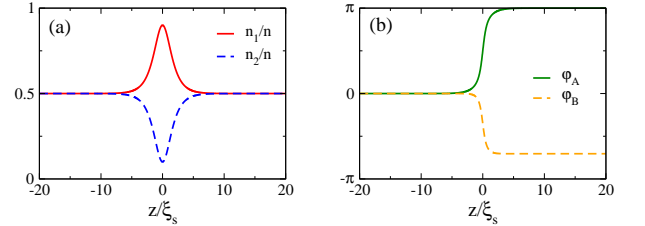}
\caption{Profiles of a magnetic soliton with velocity $%
V/c_s=0.6$. (a) The red solid and blue dashed lines are the density distributions of the
two components, satisfying $(n_1+n_2)/n=1$. (b) The green solid and yellow dashed lines are
the relative and total phases as a function of the coordinate.}
\label{fig:soliton}
\end{figure}

\begin{figure}[b!]
\includegraphics[width=8.0cm]{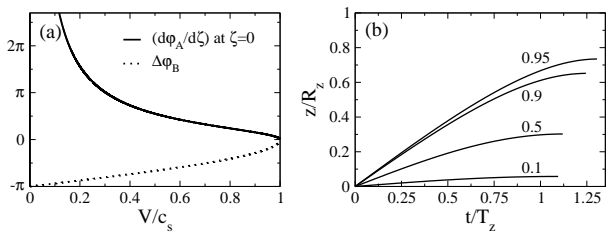}
\caption{(a) Velocity dependence of the soliton phases in a uniform system.
The solid and dotted lines are the slope of the relative phase
$d\protect\varphi_A/d\protect\zeta$ at $\protect\zeta=0$ and the increment of
the total phase $\Delta\varphi_B=\varphi_B(\zeta=+\infty)-\varphi_B(\zeta=-\infty)=-2\arccos U$.
(b) Time dependence of the magnetic soliton position in a harmonic trapping potential.
The four lines represent the motion of the soliton from the trap center $z=0$ to the
turning position for different initial velocities ($U_0=0.1,0.5,0.9,0.95$).}
\label{fig:dependence}
\end{figure}

A typical example of the soliton density distribution is shown in Fig.~\ref%
{fig:soliton}(a) for the positive value $U=V/c_s=0.6$ of the velocity. The
figure shows that the spin polarization vanishes at large distance from the
soliton. Since $0\leq |U|\leq 1$, the velocity of the magnetic soliton cannot
exceed the spin sound velocity $c_{s}$. The magnetization $(n_1-n_2)/n$ in the
center of the soliton is given by $m_{0}=\sqrt{1-U^{2}}$. It reaches its maximal
value one for the static solution ($U=0$), while vanishes as $V$ approaches the
spin sound velocity. In this latter limit the magnetic soliton behaves like a
spin wave packet. The width of the magnetic soliton, fixed by the spin healing
length $\xi_s$, is amplified by the factor $1/\sqrt{1-U^2}$. Despite the fact
that the central magnetization and the width of the soliton depend on its velocity,
the total magnetization, defined by $\int_{-\infty}^{+\infty} dz (n_1-n_2)/n$ is
velocity independent and given by the analytic result $\pi \xi_s$.
Figure~\ref{fig:soliton}(b) illustrates the phases of the magnetic soliton for $U=0.6$,
showing that $\varphi_A$ exhibits an exact $\pi$ phase jump. For negative velocities,
the phase jumps of $\varphi _{A}$ and $\varphi _{B}$ will change sign accordingly.
Figure~\ref{fig:dependence}(a) characterizes the relative and total phases of the
magnetic solitons. Although the relative phase $\varphi_A$ always has a $\pi 
$ phase jump according to Eq.~(\ref{Eq-phiA}), its slope at the
soliton center ($\partial_{\zeta }\varphi_{A}|_{\zeta =0}$) is steeper for
solitons with a slower velocity and becomes a step function for the static magnetic
soliton. Differently from $\varphi_{A}$, the asymptotic phase jump of $\varphi _{B}$
instead varies as a function of the velocity and $\varphi _{B}$ becomes zero for
$U\rightarrow 1$.

\textit{Energy and effective mass.---}Solitons can usually be described as
quasiparticles where the energy plays the role of the Hamiltonian. The
importance of the energy also relies on the fact that it is a useful
quantity to study the dynamics of solitons in harmonic traps\cite%
{Busch2000,Konotop2004}. In the case of uniform systems (no external
trapping potential) the energy of the soliton can be evaluated
straightforwardly as the difference between the grand canonical energies in
the presence and in the absence of the soliton~(see~\cite{Pitaevskii2016},
Chap.5). We find $\epsilon =n\hbar c_{s}\sqrt{1-V^{2}/c_{s}^{2}}$, which is
maximal for the static soliton and vanishes when $V=c_{s}$. For
small values of velocities, the soliton behaves like a quasi-particle with a
negative effective mass $m_{eff}=-n\hbar /c_{s}$. Solitons with negative
effective mass are not stable against nonuniform transverse snake
fluctuations. When the transverse size is larger than the width of the
soliton, the soliton decays into vortices as we will discuss in the
following.

Using the energy $\epsilon$, one can justify our main assumption that the total
density is unaffected by the presence of a magnetic soliton. Let us consider a
static soliton. Then $\epsilon =\hbar n^{3/2}\sqrt{\delta g/2m}$ and one
can calculate the depletion of number of atoms in the soliton using the
thermodynamic relation $N_{D}\equiv \int_{-\infty }^{\infty }\left[ n(z)-n%
\right] dz=-\partial \epsilon /\partial \mu ,$ where $\mu =ng$ is the
chemical potential. A simple calculation gives $N_{D}=-3n\xi _{s}(\delta g/2g)$%
. One can estimate the density perturbation near the soliton center as $\left\vert
n(z)-n\right\vert \sim \left\vert N_{D}\right\vert /\xi _{s}\sim n\delta
g/g\ll n$.

\textit{In-trap oscillations.---}The particle-like nature of solitons is
revealed by its long time stable oscillations in a 1D or elongated 2D
harmonic traps with axial trapping frequency equal to $\omega_z$. The
oscillation period of dark solitons of a single component Bose gas in a
harmonic trap is $\sqrt{2}T_{z}$, where $T_{z}=2\pi/\omega_z$ is the
oscillator period~\cite{Busch2000,Konotop2004}. For the dark-bright solitons
studied in Ref.~\cite{Busch2001}, the oscillation is much slower and depends
on the population of the bright-soliton component.

To describe the oscillation of magnetic solitons in a harmonic trap, we
consider the energy of the soliton in the local density approximation $%
\epsilon(z)=n(z)\hbar \sqrt{c_{s}^{2}(z)-V^{2}(z)}$, where $n(z)$, $c_{s}(z)$
and $V(z)$ are the total density, the spin sound velocity and the velocity
of the magnetic soliton at position $z$. Let us assume that a magnetic
soliton is initially engineered at $z=0$ with energy $\epsilon_{0}=n_{0}%
\hbar \sqrt{c_{s0}^{2}-V_{0}^{2}}$ where $n_{0}$, $c_{s0}$ and $V_{0}=U_0
c_{s0}$ are the density, spin sound velocity and the soliton velocity at the
trap center. Since the energy of the soliton remains $\epsilon_0$ in the
following evolution, the velocity is therefore evaluated as 
\begin{equation}
V(z)=\frac{dz}{dt}=\sqrt{\frac{n(z)\delta g}{2m}-\frac{\epsilon_0^2}{%
n^2(z)\hbar^2}},  \label{Eq-velocity}
\end{equation}%
where the total density at the position of the soliton can be approximated
by $n(z)=n_{0}(1-z^{2}/R_{z}^{2})$, the Thomas-Fermi radius $R_{z}$ of the
condensate being given by $m\omega _{z}^{2}R_{z}^{2}/2=n_{0}g$.

\begin{figure}[tbp]
\includegraphics[width=8.6cm]{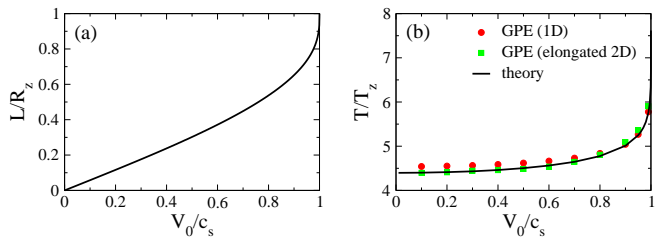}
\caption{In-trap oscillation of a magnetic soliton. (a) The
turning position $L/R_z$ is plotted as a function of the initial velocity $%
V_0/c_s$ of the magnetic soliton created at the center of the harmonic trap.
(b) The black line is the theoretical prediction of the scaled oscillation
period $T/T_z$. The red circles and green squares are the data from the
numerical solution of GPE for a trapped gas of $^{23}$Na atoms in a 1D or an
elongated 2D geometry with chemical potential $\protect\mu=h\times 1210 Hz$ and
longitudinal trapping frequency $\protect\omega_z=2\protect\pi\times 13
Hz $. The transverse frequency for the 2D system is $\protect\omega_y=2\protect\pi\times 130 Hz$,
thus $\protect\mu=9.3\protect\omega_y$. The $s$-wave scattering lengths
are $a_{11}=a_{22}=54.54a_0$, $a_{12}=50.78a_0$ with $a_0$ the Bohr radius.}
\label{fig:frequency}
\end{figure}

The oscillation of the magnetic soliton can be explored by solving Eq.~(\ref{Eq-velocity}) with the initial condition $z(t=0)=0$. Figure~\ref{fig:dependence}(b) shows the trajectory of the soliton moving from $z=0$ to the turning point $z=L$ in the interval $t=T/4$, where $T$ is the period of a complete oscillation. The turning point, which determines the amplitude of the soliton oscillation, can be explicitly calculated by the position where the velocity vanishes $V(z=L)=0$. Using the relation $\epsilon(L)=\epsilon_{0}$, we have $L/R_{z}=\sqrt{1-(1-U_{0}^{2})^{1/3}} $ which is plotted in Fig.~\ref{fig:frequency}(a) as a function of the initial velocity. Direct integration of Eq.~(\ref{Eq-velocity}) from $z=0$ to $z=L$ gives the result 
\begin{equation}
\frac{T}{T_{z}}=\frac{4}{\pi }\sqrt{\frac{g}{\delta g}}\int_{0}^{L/R_z}\frac{v(\beta )d\beta }{\sqrt{v^{3}(\beta )-1+U_{0}^{2}}},  \label{Eq-Freq}
\end{equation}
for the period of the soliton oscillation, with $v(\beta )=1-\beta ^{2}$ and $\beta=z/R_z$.
The integral in Eq.~(\ref{Eq-Freq}) can be solved analytically in two
extreme cases: \text{(i)} The slowly moving soliton. For $U_{0}\rightarrow 0$, the above integral
can be solved with the help of Taylor expansions. Calculation gives the
turning position as $L/R_{z}=U_{0}/\sqrt{3}$ and the oscillation period
reads $T/T_{z}=2\sqrt{g/3\delta g}$, showing
that, in contrast to the usual dark solitons, the oscillation period of the
magnetic soliton depends on the interaction parameters.
\text{(ii)} The other interesting limiting case is for $U_{0}\rightarrow 1$,
that is when the velocity of the magnetic soliton approaches the spin sound
velocity. In this case $L/R_{z}\rightarrow 1$ and we find $T/T_{z}=2\sqrt{%
g/\delta g}$, which is $\sqrt{3}$ times larger than that in the slow moving
limit.We emphasize that in this extreme case, the magnetic soliton will
reach the edge of the Bose gas where the above approximation is no longer
applicable~\cite{Konotop2004}.

For other intermediate values of initial velocities, the oscillation period
is obtained by a direct numerical integration of Eq.~(\ref{Eq-Freq}) and the
result is shown in Fig.~\ref{fig:frequency}(b). Here we have used the
typical values of the scattering lengths of $^{23}$Na where $\delta
g/g\approx 0.07$. The oscillation period for small velocity is about $4.4$
times of $T_{z}$ which is much larger than the period $T/T_{z}=\sqrt{2}$ of
a dark soliton in a single component Bose gas ~\cite{Busch2000,Konotop2004}.
The above prediction for the oscillation period of the magnetic soliton very
well agrees with the numerical simulation of the GPE for a 1D condensate, as
well as for a highly elongated 2D configuration.

\begin{figure}[t!]
\includegraphics[width=5.5cm,height=3.0cm]{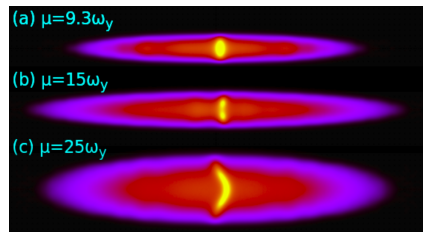}
\caption{Stability of magnetic solitons in 2D. The density
distributions $n_1$ are shown for the following chemical
potentials and aspect ratios $\protect\lambda=R_z/R_y$: (a) $\protect\mu=9.3%
\protect\omega_y$, $\protect\lambda=10$; (b) $\protect\mu=15\protect\omega_y$%
, $\protect\lambda=10$; (c) $\protect\mu=25\protect\omega_y$, $\protect%
\lambda=5$. In each case, the soliton is imprinted at the center of the
Bose gas with an initial velocity of $V_0=0.1c_s$ along the horizontal $z $
direction. The other parameters are the same as in Fig.~\protect\ref{fig:frequency}.}
\label{fig:stability}
\end{figure}

\textit{Stability in two dimensions.---}For a 2D system, when the transverse size of the
condensate is larger than the width of the magnetic solitons, the solitons
become unstable and start to bend while moving due to the snake instability.
Since the width of magnetic solitons increases with their velocity, fast
moving solitons are more stable than slow moving solitons. To showcase the
instability of the magnetic solitons at low velocities, we plot the 2D
density distributions of the first component $n_1$ in a harmonic trap obtained from
the numerical simulation of the GPE. The soliton is imprinted at the trap
center at $t=0$ with an initial velocity $V=0.1c_{s}$ and imaged after a
short evolution time $t=29$ms. As shown in Fig.~\ref{fig:stability}, with
the increase of the chemical potential the transverse size of the condensate
increases. For $\mu=9.3\omega _{y}$, where $\omega _{y}$ is the transverse
trapping frequency, the magnetic soliton is stable and oscillates in trap in
good agreement with the 1D result. For $\mu =15\omega _{y}$, the magnetic
soliton still moves and oscillates in the trap. However, vortex pairs soon
appear in the second component and the system alternatively oscillates
between a vortex pair and a magnetic soliton. For a 3D elongated harmonic
trap, vortex rings are expected to appear and one consequently expects to
observe an oscillating evolution between vortex ring and magnetic soliton
configurations, in analogy with the oscillation between a vortex ring and a
dark soliton already observed in a one-component Bose gas \cite{Shomroni2009}%
. For $\mu =25\omega _{y}$, the magnetic soliton is unstable, quickly starts
to bend and eventually decays into vortices.

\textit{Conclusion.---}We have investigated a magnetic
soliton moving in a spinor Bose gas. The properties of the magnetic solitons
and their oscillation dynamics in a harmonic trap are
characterized and predicted analytically and numerically. We believe that
our theoretical predictions will stimulate new experimental work in the
field of spinor condensates. Since the relative phase of each magnetic
soliton is $\pi $, a pair of solitons with opposite magnetization and moving
in opposite directions could be engineered by imprinting a $2\pi $ relative
phase between the two components, generating a domain wall of
tunable width \cite{Son2002}. Other important questions, such as the
stability analysis of magnetic solitons in higher dimensions, remain to be
investigated in the future.

\begin{acknowledgements}
We thank G. Ferrari, G. Lamporesi, F. Dalfovo, A. Recati and M. Tylutki for stimulating
discussions. This work was supported by ERC through the QGBE grant, by the
QUIC grant of the Horizon2020 FET program and by Provincia Autonoma di
Trento.
\end{acknowledgements}


\begin{thebibliography}{99}
\bibitem{RbK} In the case of $^{41}$K and $^{87}$Rb, one should instead
consider the states $|F=1; m_F=-1\rangle$ and $|F=1; m_F=0\rangle$ in order
to ensure the positiveness of $\delta g$. In this case $g_{11} \approx
g_{22} $ and $\delta g/g \approx 0.013$ and $0.0025$ for the two atomic species,
respectively.
\end{thebibliography}
\end{document}